\documentclass[11pt,A4]{article}

\usepackage{amsmath}
\usepackage{amsfonts}
\usepackage{amssymb}
\usepackage{graphicx}
\usepackage{cite}
\usepackage[pdftex]{hyperref}

\newlength{\xtrawidth}
\newlength{\xtraheight}
\newcommand{\beq}{\begin{equation}}
\newcommand{\eeq}{\end{equation}}
\newcommand{\bea}{\begin{eqnarray}}
\newcommand{\eea}{\end{eqnarray}}

\newcommand{\IZ}{\mathbb{Z}}
\newcommand{\IP}{\mathbb{P}}
\newcommand{\CN}{{\mathcal N}}
\newcommand{\eref}[1]{(\ref{#1})}
\newcommand{\comment}[1]{}
\newcommand{\mycaption}[1]{\caption{{\sf {\small #1 }}}}

\begin{document}

\title{\Huge An Abundance of Heterotic Vacua\\}
\author{Maxime Gabella$^1$\footnote{gabella@physics.ox.ac.uk}~,
Yang-Hui He$^{1,2}$\footnote{hey@maths.ox.ac.uk}~
and Andre Lukas$^1$\footnote{lukas@physics.ox.ac.uk}}
\date{}
\maketitle
\begin{center}
\vskip -1cm
{
{\it $~^1$  Rudolf Peierls Centre for Theoretical Physics, \\
Oxford University, 1 Keble Road, \\
Oxford,  OX1 3NP, U.K.}\\
{\it $~^2$ Merton College, Oxford, OX1 4JD, U.K.}
}
\end{center}

\begin{abstract}
We explicitly construct the largest dataset to date of heterotic vacua arising
from stable vector bundles on Calabi-Yau threefolds. 
Focusing on elliptically fibered Calabi-Yau manifolds with spectral cover bundles, we show that the number of heterotic models with non-zero number of generations is finite.
We classify these models according to the complex base of their Calabi-Yau threefold and to the unification gauge group that they preserve in four dimensions.
This database of the order of $10^7$ models, which includes potential Standard Model
candidates, is subjected to some preliminary statistical analyses.
The additional constraint that there should be three net generations of particles gives
a dramatic reduction of the number of vacua. 
\end{abstract}

\newpage

\tableofcontents

\section{Introduction}

Heterotic string and M-theory remain promising approaches toward building
phenomenologically realistic models of four-dimensional particle physics. 
Since the beginning of superstring phenomenology in refs.~\cite{Candelas:1985en,Witten:1985bz} two decades ago,
much progress has been made. It is by now well-established that
compactification of the $E_8\times E_8$ heterotic string on Calabi-Yau threefolds
endowed with stable holomorphic $SU(n)$ vector bundles
leads to supersymmetric Grand Unified Theories  (GUT) in four dimensions. 
Furthermore, introducing
Wilson lines can break the GUT gauge group down to the Standard Model (SM) group.

The advantage of this method is that very precise and succinct mathematical
quantities, namely cohomology groups of the vector bundles, encode the
particle spectrum and interactions of the four-dimensional physics.
A considerable amount of work has been devoted to developing techniques
for constructing vector bundles and computing the associated
cohomology groups   
\cite{Donagi:2004ia,Braun:2005ux,Bouchard:2005ag,Donagi:2004su,Distler:2007av,Blumenhagen:2006wj,Bouchard:2008bg,Distler:1987ee,Kachru:1995em,Blumenhagen:1997vt,Andreas:2007ev,Anderson:2007nc,Anderson:2008uw,Lukas:1998hk,Donagi:1999gc,Grassi:2000fk,Andreas:1998ei,Curio:1998vu}.

Much of the literature has focused on finding specific
vector bundles on select threefolds which lead to realistic
theories. However, given the multitudes of Calabi-Yau threefolds
and of potential vector bundles on them, it is important to have a view of the
global picture, analyze the space of models, and especially determine how many
models are quasi-realistic. Only recently has there been an effort to understand
this heterotic landscape. In ref.~\cite{Candelas:2007ac}, a special corner
has been found which tends to produce realistic models
(see ref.~\cite{Gmeiner:2005vz,Gmeiner:2007zz,Dienes:2006ut}).

Indeed, with the advances of computing power and software in computational
algebraic geometry, a novel perspective on heterotic compactification has
been proposed in ref.~\cite{Anderson:2007nc}.
In ref.~\cite{Anderson:2008uw},
so-called monad bundles are constructed over a large dataset of Calabi-Yau threefolds known as CICYs, or
Complete Intersection Calabi-Yau manifolds \cite{Candelas:1987kf}.
One of the advantages of this set is the embedding into a projective ambient space which facilitates
standard techniques for calculating the requisite bundle cohomology groups.
However, proving stability in general for these bundles requires separate
treatment and is rather difficult; a comprehensive procedure is still elusive \cite{stabpaper}.

Luckily, a systematic technique for creating stable vector bundles
does exist for a wide class of Calabi-Yau manifolds, constructed by elliptic fibration \cite{Ron,FMW}.
These elliptically fibered
threefolds are tori fibered over a complex base surface which also
have a zero section. Such manifolds have been completely classified 
\cite{Gross:1994,Morrison:1996na}. 
Stable vector bundles can be conveniently
constructed over them using the spectral cover method  \cite{Ron,FMW}.
An initial attempt at classifying such vector bundles over this dataset of
Calabi-Yau threefolds was undertaken in ref.~\cite{Donagi:2004ia}.

The purpose of the current paper is to classify, as much as computer
power allows, the spectral cover bundles over elliptically fibered threefolds
and to examine some of their properties and statistical features in the light of
basic physical constraints, such as the three-generation constraint. It turns out
that a tremendous number is readily found. This constitutes the largest explicit 
dataset of stable vector bundles to date. For the Calabi-Yau manifolds constructed from Hirzebruch base spaces we find about $50,000$ inequivalent cases. After imposing basic physical constraints this number is drastically reduced to about $1,700$.  For the Calabi-Yau manifolds based on del Pezzo surfaces, $d\mathbb{P}_r$, we find that the number of consistent vacua increases dramatically with $r$, and we are able to perform a complete classification for $r\leq 3$ only. Particularly, for $r=3$ we find over $11$ million models, of which about $400,000$ are still compatible with basic physical constraints. For Calabi-Yau manifolds based on higher del Pezzo surfaces, the classification is limited by computational power. We perform some statistical analysis of the models and discuss model-building prospects.

The organization of the paper is as follows.
In Section 2, we lay out the mathematical construction of stable vector bundles over
elliptically fibered Calabi-Yau threefolds and review how physical aspects such as the number of generations, GUT gauge groups, and  anomaly cancellation are conveniently encoded. 
We summarize all requisite physical constraints as explicit Diophantine inequalities in Section 3 and
show that the number of solutions is finite. In Section 4, we proceed to classify stable $SU(n)$ 
vector bundles from the spectral cover construction on the elliptically fibered Calabi-Yau threefolds.
We conclude with discussions and prospects in Section 5.

\section{Heterotic Compactification}
Let us begin with a brief review of the compactification of heterotic string theory
and, in the non-perturbative regime, heterotic M-theory. This section serves as a reminder of the
mathematical constructions and physical constraints
involved. We begin by motivating the need for stable holomorphic vector bundles on Calabi-Yau threefolds
and then specialize to a wide class of elliptically fibered Calabi-Yau threefolds. We then briefly review the spectral cover method for constructing stable bundles on these manifolds.

\subsection{Stable Bundles on Calabi-Yau Threefolds}
The ``traditional" way to relate the ten-dimensional string theory to a four-dimensional
space-time $M^4$ is to start with the $E_8\times E_8$ heterotic string theory on a background 
\beq
M^{10} = M^4 \times X \ ,
\eeq
where $X$ is a compact six-dimensional manifold (for a review see ref.~\cite{Green:1987mn}). 
Historically, this was the first approach toward string phenomenology~\cite{Candelas:1985en,Witten:1985bz}.

The requirement of unbroken $\mathcal{N}=1$ supersymmetry in four dimensions further specifies the construction. The standard solution is that
the non-compact (and maximally symmetric) space-time $M^4$ is flat Minkowski space while the compact manifold $X$ is a Calabi-Yau threefold, that is, a complex K\"ahler manifold with a metric of $SU(3)$ holonomy.
Equivalently, by Yau's theorem, $X$ is complex K\"ahler and has vanishing first Chern class of its tangent bundle, $c_1(X)=0$.

It is common to declare one of the $E_8$ gauge groups to be the ``visible sector,'' which is to contain the particles of
the SM, and the other $E_8$ to be the ``hidden sector.''  Additional hidden sectors can arise from the world-volumes of five-branes which may be included in the compactification. Throughout this paper we will not consider these hidden sectors explicitly but we will ensure that a choice of data in the hidden sectors which leads to an overall  consistent model exists. 
In order to reduce the visible $E_8$ group we allow a vector bundle $V$ on $X$ with gauge group $G$. The low-energy gauge group $H $ will then be the commutant of $G$ in $E_8$.  Moreover, we will focus on the cases $G=SU(n)$ for $n=3,4,5$, so that $H$ is one of the standard grand unification groups $E_6$, $SO(10)$, and $SU(5)$, respectively. Subsequently, one needs to introduce appropriate Wilson lines in order to break $H$ to the Standard Model gauge group.
Because $G$ is a special unitary group the vector bundle $V$ satisfies the condition
\beq
c_1(V) = 0\; .
\eeq

In order to preserve $\CN=1$ supersymmetry, the gauge connection $F$ on $X$ must satisfy the 
hermitian Yang-Mills equations $F^{mn} = F^{\bar{m}\bar{n}} = g_{m\bar{n}}F^{m\bar{n}} = 0$.
These are rather difficult equations to solve. Luckily,
theorems by Donaldson, Uhlenbeck and Yau \cite{duy}
state that a holomorphic vector bundle $V$ on $X$ will admit such a connection if and only if it is \makebox{(poly-)stable}.
A complicated set of partial differential equations is thereby translated to a problem of pure algebraic geometry.
The proof of stability is still a difficult issue (see for example refs.~\cite{Anderson:2007nc,Anderson:2008uw,stabpaper}), but fortunately the spectral cover construction used in the present paper will automatically guarantee this.

Now, in order to be consistent at the quantum level, we need to impose Green-Schwarz anomaly cancellation. This translates to a constraint on the second Chern classes of the tangent bundle of the compact manifold $X$, the visible sector bundle $V$, the hidden bundle $\tilde{V}$,  and the holomorphic curve $W$ wrapped by five-branes~\cite{Horava:1995qa}.
This constraint reads (see for example \cite{Lukas:1998hk,Donagi:1999gc})
\beq\label{CohConstrW0}
c_2(X)-c_2(V)-c_2(\tilde{V})=[W],
\eeq
where $[W]$ is the homology class of $W$ and provided that both $V$ and $\tilde{V}$ are vector bundles with vanishing first Chern class. Since $W$ is a holomorphic curve, its associated  class $[W]$ is effective, or, in other words, it is an element of the Mori cone of $X$. Given a Calabi-Yau manifold $X$ and a visible bundle $V$, a simple way to make sure that the anomaly condition can be satisfied is to demand that
\beq\label{CohConstrW}
 c_2(X)-c_2(V)\in\mbox{ Mori cone of }X\; .
\eeq
In this case, one can always find a five-brane curve $W$ such that the anomaly condition~\eqref{CohConstrW0} is indeed satisfied for a trivial bundle $\tilde{V}$ (although there may well be alternative choices which involve a non-trivial hidden bundle $\tilde{V}$). We will henceforth use the condition \eref{CohConstrW} for the purpose of classifying spectral cover bundles.

\subsection{Elliptic Fibration}
As discussed above, the first object we need is a Calabi-Yau threefold $X$. 
In this paper, we will focus on the rich data set of elliptically fibered Calabi-Yau threefolds since they allow for a convenient construction of stable vector bundles.

An {\em elliptically fibered Calabi-Yau threefold} $X$ is defined by a fibration
\beq
X \stackrel{\pi}{\to} B
\eeq
over a complex base surface $B$ such that the fiber $\pi^{-1}(b)$ is an elliptic curve for each generic point $b\in B$.
We are referring to an elliptic curve rather than a torus because we require the existence of a global section
\beq
\sigma: B\to X,
\eeq
which associates to every point in $B$ the zero element of the addition law on the elliptic curve.

The existence of a global section is a surprisingly strong constraint~\cite{Gross:1994}, and as a result the complex base surface can only be one of the following \cite{Morrison:1996na}~: Hirzebruch surfaces and their blow-ups, del Pezzo surfaces, and Enriques surfaces. We will introduce these surfaces in detail in Section 4.

One advantage of this fibered construction is that the Chern classes of $X$ can be easily
expressed in terms of those of the base surface $B$ as \cite{FMW}
\bea\label{c2X}
c_1(X) &=& 0, \\
c_2(X) &=& c_2(B) + 11 c_1(B)^2 + 12 \sigma c_1(B),\\
c_3(X) &=& -60 c_1(B)^2 \ . \label{c3X}
\eea
These formulae will be useful later. We remark that the Euler number of $X$ is simply
\beq\label{eulerB}
\chi(X) = \int_X c_3(X) = -60 c_1(B)^2 \ .
\eeq

\subsection{Spectral Cover Construction}\label{s:spec}
As mentioned earlier, having background gauge fields which satisfy the hermitian Yang-Mills 
equations is equivalent to choosing a stable holomorphic vector bundle.
The greatest advantage of elliptically fibered Calabi-Yau threefolds is that a systematic and relatively
straightforward method has been developed to construct holomorphic vector bundles on them 
which are guaranteed to be stable.
This is the so-called {\em spectral cover construction} \cite{Ron,FMW}.

The idea is to first construct the bundles on individual elliptic fibers using a classic result on stable bundles over elliptic curves due to Atiyah. Then these bundles over individual fibers are patched together over the base.
In summary, an $SU(n)$ bundle $V$ over $X$ is given by the spectral data, consisting of the following two pieces~:
\begin{itemize}
\item The {\em spectral cover} $\mathcal{C}_V$~:  this is an $n$-fold cover of the base and
	is thus a divisor (a linear combination of hypersurfaces) in $X$ with degree $n$ over $B$.
	This implies that the cohomology class of $\mathcal{C}_V$ in $H^2(X,\mathbb{Z})\simeq H_4(X,\IZ)$ is of the general form
	\beq
	[\mathcal{C}_V] =  n\ \sigma +  \eta \ ,
	\eeq
	where $\sigma$ is the class of the zero section, and $\eta$ is a curve class in $H^2(B,\mathbb{Z})$.
	The class $\eta$ must be effective in $B$, which means that it must be possible to express 
	it as a linear combination of effective classes $S_i\in H^2(B,\mathbb{Z})$ with non-negative coefficients~:
	\beq
	\eta = \sum_i a_i  S_i, \qquad \text{with} \qquad a_i\ge 0.	
	\eeq
	The subset of effective classes forms a cone in $H_2(B,\mathbb{Z})$ called the {\em Mori cone}.
\item The {\em spectral line bundle} $\mathcal{N}_V$~: this is a line bundle on $\mathcal{C}_V$ with first Chern class
\begin{equation}
\label{c1N}
c_{1}(\CN_V)=n(\frac{1}{2}+\lambda)\sigma+(\frac{1}{2}-\lambda)
\pi^{*}\eta+(\frac{1}{2}+n\lambda)\pi^{*}c_{1}(B) \ .
\end{equation}
The parameter $\lambda$ has to be either integer or half-integer depending on the rank $n$ of the $SU(n)$ structure group~:
\begin{equation}
\lambda =  \left\{
\begin{array}{cl}
	m+1/2 & \text{ if $n$ is odd}, \\
	m & \text{ if $n$ is even},
\end{array} \right.
\end{equation}
where $m\in\mathbb{Z}$.
When $n$ is even, we must also impose $\eta = c_1(B) \textrm{ mod } 2$, 
by which we mean that $\eta$ and $c_1(B)$ differ only by an even element of $H^2(B,\mathbb{Z})$.
\end{itemize}

The holomorphic $SU(n)$ vector bundle $V$ on $X$ can be extracted from the above data by a so-called
Fourier-Mukai transformation~: $(\mathcal{C}_V,\mathcal{N}_V) \,\stackrel{FM}{\longleftrightarrow}\, V$
(see refs.~\cite{Andreas:2004uf,Donagi:2008ca} 
for some applications of this transformation in string theory).
The Chern classes of $V$ are given in terms of the spectral data as
\cite{FMW,Curio:1998vu,Andreas:1998ei}
\bea
   c_1(V) &=& 0, \label{c1} \\
   c_2(V) &=& \eta \sigma - \frac{n^3 - n}{24} c_1(B)^2
              + \frac{n}{2} \left(\lambda^2 - \frac 1 4\right) \eta \cdot
                 \left(\eta - nc_1(B)\right),  \label{c2V} \\
   c_3(V) &=& 2 \lambda \sigma \eta \cdot \left(\eta - nc_1(B) \right). \label{c3V}
\eea

One of the advantages of the spectral cover construction is that stability of $V$ can be guaranteed by
fairly simple algebraic conditions~:
the vector bundle $V$ is stable if $\mathcal{C}_V$ is irreducible.
This will be the case if we impose the conditions (see for example \cite{Donagi:2004ia})
\bea\label{base-pt}
&& \text{the linear system $|\eta|$ is base-point free in $B$,}\\
&& \text{$\eta - n c_1(B)$ is an effective curve in $B$.}\label{stab2}
\eea
We recall that
the {\em linear system} $|\eta|$ is the set of all effective curves linearly equivalent to $\eta$ 
(that is, which only differ from $\eta$ by the divisor of a meromorphic function \cite{hart}).
It is {\em base-point free} if its members have no common intersection.
We will make these two rather technical conditions more explicit for the surfaces we will encounter in Section 4.

Finally, the five-brane class $W$ can be split up into a curve class $W_B$ in the base surface $B$
and the fiber class $F$ of the elliptic fibration, so that
\beq 
W = W_B +a_f F,
\eeq
with $a_f$ some integer.
For most of the base spaces that we will consider, the class $W$ is effective if and only if the 
following conditions hold~:
\bea\label{W_B}
&& W_B \,\text{ is effective},\\
&& a_f\ge0.\label{a_f}
\eea
There is an exception to this rule for Hirzebruch surfaces $\mathbb{F}_r$ with $r\ge3$ \cite{Donagi:1999gc}; 
we will discuss this exception in Section \ref{s:Fr} and properly incorporate it into our classification.

We can simplify the expressions in eq.~\eqref{W_B}.
Using eqs.~\eqref{CohConstrW}, \eqref{c2X}, and \eqref{c2V}, we can write $W_B$ and $a_f$ in terms of the 
cohomology classes of the base $B$ as
\bea
W_B &=& 12 c_1(B) - \eta, \\
a_f &=& c_{2}(B)
   + \left(11+\frac{n^{3}-n}{24}\right) c_{1}(B)^{2}
   - \frac{n}{2}\left(\lambda^{2}-\frac{1}{4}\right)
   \eta \cdot (\eta - n \ c_1(B)).
\eea

\subsection{Number of Generations}
\comment{
The compactification of the strongly coupled heterotic string is specified by choosing the Calabi-Yau threefold $X$, the vector bundle $V$, and the
structure group $SU(n)$.
The features of the low-energy theory can then in principle be computed.
The simplest quantity to compute is the net number of fermion generations $N_{\textrm {gen}}$.
}
A salient feature of heterotic compactification is that the low-energy particles are given in terms of the
vector bundle cohomology groups for $V$ \cite{Green:1987mn}; these are well-defined mathematical quantities to compute. 
In particular, for $SU(n)$ bundles, we can count the net number of generations, $N_{\textrm {gen}}'$, in the resulting Grand Unified Theory. This is a topological number and from the index theorem it can be expressed as
\bea
N_{\textrm {gen}}' 
= \frac{1}{2}\Big\vert\int_X c_3(V)\Big\vert.
\eea

If we wish to further break the grand unified group to the Standard Model group, we normally have to quotient the Calabi-Yau manifolds by a freely-acting discrete symmetry to obtain a non simply connected space and then turn on Wilson
lines. Typically the symmetry group is a cyclic group $\mathbb{Z}_k$ or a product thereof (for a recent discussion on potentially large discrete symmetries, see ref.~\cite{Doran:2007jw}). Let the order of this group be $k$. The net number of generations on the quotient manifold, $N_{\textrm gen}$, is then reduced by the order of this group and given by
\beq
N_{\textrm {gen}} =  N_{\textrm {gen}}'/k \ . \label{NN}
\eeq
For elliptically fibered threefolds it is usually not easy to find freely acting discrete symmetries and we will not explicitly attempt this in the present paper. Instead, we will use some basic necessary conditions for the existence of such a symmetry.
First of all, eq.~\eqref{NN} implies that the ``upstairs" number of generations, $N_{\textrm {gen}}'$, must be a multiple of three,
\beq\label{3k}
N_{\textrm {gen}}'= 3k \; ,
\eeq
so that the order of a discrete symmetry group which leads to three generations ``downstairs" is given by $k=N_{\textrm {gen}}'/3$. For the Calabi-Yau manifold $X$ to allow for such a discrete symmetry its Euler number must, of course, be divisible by the order $k$, so
\beq\label{euler}
\chi(X)/k \in \mathbb{N}\; .
\eeq
Eqs.~\eqref{3k} and \eqref{euler} are the two basic physical constraints which we will impose on the models found in this paper.
For practical calculations, they can be expressed in terms of the base surface and the spectral data by using eqs.~\eqref{c3V} and \eqref{c3X}. 

\section{Summary of Constraints and Finiteness}
In the previous section we have presented the rudiments of constructing stable, holomorphic $SU(n)$ vector
bundles on an elliptically fibered Calabi-Yau threefold. The requirement of anomaly
cancellation for a consistent heterotic vacuum and the physical condition of three net generations of low-energy particles 
lead to a set of constraints on these bundles. It is expedient to summarize these, now phrased in a succinct mathematical manner. In the following section, we will show how these constraints lead to a classification problem.
For recent related work, the reader is also referred to refs.~\cite{Distler:2007av} and \cite{Bouchard:2008bg}.
Combining eqs.~\eqref{base-pt}, \eqref{stab2}, \eqref{W_B}, \eqref{a_f}, \comment{\eqref{a_fHirz},} \eqref{3k}, and  \eqref{euler}, we gather the six following constraints.
\begin{itemize}

\item {Stability of the vector bundle $V$~:}

\begin{enumerate}
\item[1.)] $|\eta|$ \, must be base-point free.
\item[2.)] $\eta - n c_1(B)$ \, must be effective.
\end{enumerate}

\item {Anomaly cancellation with five-branes (effectiveness of $W$)~:}

\begin{enumerate}
	\item[3.)] $W_B = 12 c_1(B) - \eta$ \, must be effective.
	\item[4.)] $a_f = c_{2}(B)
   + \left(11+\frac{n^{3}-n}{24}\right) c_{1}(B)^{2}
   - \frac{n}{2}\left(\lambda^{2}-\frac{1}{4}\right)
   \eta \cdot (\eta - n \ c_1(B)) \ge 0$,\\
where\footnote{With an exception for $\mathbb{F}_{r\ge 3}$, in which case $a_f \ge 96+a r-2a-2b$, see Section 4.1.} $\lambda = m+1/2$ if $n$ is odd, and
$\lambda = m$ if $n$ is even, with $m\in\mathbb{Z}$.

\end{enumerate}

\item {Three generations~:}
\begin{enumerate}
\item[5.)] $N_{\textrm {gen}}'=\vert\lambda \ \eta \cdot (\eta - n \ c_1(B))\vert = 3k$, \quad
with $k\in\mathbb{N}$.
\item[6.)] $k$ divides $\chi(X)$~: \quad $60 c_1(B)^2 /k \in \mathbb{N}$.
\end{enumerate}

\end{itemize}
We recall that the rank $n$ of the structure group $SU(n)$ equals 3, 4 or 5, corresponding respectively to low-energy gauge groups $E_6$, $SO(10)$, or $SU(5)$. Note that all the constraints are conveniently expressed in terms of quantities on 
the base surface $B$. The Chern classes can easily be computed for the various allowed base surfaces, and the curve $\eta$ can be expanded into a basis of second homology.  A solution to these constraints will consists of a set of 
coefficients that specify the effective class $\eta$ in terms of the generators of the Mori   
cone, as well as the value of the arbitrary integer or half-integer parameter $\lambda$. We note that the above set of conditions really splits into two logically somewhat distinct parts. Conditions 1.) to 4.) guarantee the existence of a consistent heterotic vacuum and our initial classification will, therefore, focus on these first four constraints. Constraints 5.) and 6.), on the other hand, are constraints of a ``phenomenological" nature and will only subsequently be imposed on the set of consistent vacua in order to filter out promising models. Hence, our classification problem can be stated as follows.
\begin{quote}
{\em Find all $\eta$ (specified by non-negative integer coefficients of an expansion in the basis of the Mori cone of $B$) and $\lambda$ (integer or half-integer according to the rank $n$) such that the above constraints 1.) to 4.) are satisfied. Within this set find all cases which in addition satisfy constraints 5.) and 6.).}
\end{quote}

We can immediately make some observations.
First, note that the intersection number $\eta \cdot (\eta - n c_1(B))$ appears in both
constraints 4.) and 5.). Models with a zero net number of generations are not particularly interesting, and to exclude such cases we demand, in addition to constraint 4.), that
\beq\label{cond4.1}
\lambda \ne 0, \quad \eta \cdot (\eta - n c_1(B)) \ne 0 \ .
\eeq
These will be included with constraints 1.) to 4.) in our actual initial classification. A technical reason for demanding a non-zero number of generations has to do with the issue of finiteness which we will address shortly.
On the so-obtained data set we will then impose the three-generation constraints 5.) and 6.). 

Second, in all our constraints, $\lambda$ only appears as a square or an absolute value; 
thus for every solution with positive $\lambda$ there is also a solution 
with negative $\lambda$. Since the third Chern class of $V$ depends explicitly on $\lambda$
itself, as seen from \eqref{c3V},  these two sets of solutions are actually different bundles.

Finally, let us consider the issue of whether the number of solutions is finite or infinite. Let us examine conditions 2.) and 3.). Crucially, we see that these two conditions have
opposite signs in front of $\eta$. Effectiveness is a positivity condition and this means that
2.) and 3.) provide upper and lower bounds for the coefficients in the expression of $\eta$. If the
Mori cone is finitely generated, then this implies that there is only a finite number of possible solutions of $\eta$. As we will see below, all of our base surfaces have a finite-dimensional Mori cone, except the ninth del Pezzo surface. Luckily, this particular surface will be ruled out by the requirement of a non-zero net number of generations, that is by the condition~\eqref{cond4.1}. 

With a finite possible set of solutions for $\eta$, condition 4.) constitutes a quadratic inequality
for $\lambda$ if the coefficient $\eta \cdot (\eta - n c_1(B))$ does not vanish. This is precisely what we have required in eq.~\eqref{cond4.1} in order to have a non-vanishing number of families. As a result, the number of possible $\lambda$ values is finite. This finiteness result is the technical reason for the non-vanishing condition \eqref{cond4.1}. In our detailed calculations below, it will turn out that there are some cases for which $\eta \cdot (\eta - n c_1(B))$ is indeed zero. They may lead to {\em an infinite family of stable vector bundles} satisfying the anomaly constraints, although all of them with a zero number of generations. 
We will not presently address these bundles.

Hence, since there is a finite number of solutions to our variables $\lambda$ and $\eta$,
we immediately have a nice finiteness result.
\begin{quote}
{\em There is a finite number of solutions to constraints 1.) to 4.), together with the condition~\eqref{cond4.1}. That is, there is a finite number of spectral cover $SU(n)$ vector bundles on elliptically fibered Calabi-Yau threefolds which lead to anomaly-free heterotic vacua with a non-vanishing number of generations.}
\end{quote}
From these we can select three-generation models by imposing constraints 5.) and 6.), which will be done explicitly in the ensuing section.

An initial attempt at classifying these bundles was made in ref.~\cite{Donagi:2004ia}; however, at the time, interest was more in the development of techniques of computing particle spectrum. Recently, a complete classification was achieved for positive monad bundles over Complete Intersection Calabi-Yau threefolds in refs.~\cite{Anderson:2007nc,Anderson:2008uw}, and a similar finiteness result as above was encountered.

\section{Classification of Stable Bundles}
We have laid the foundation and presented the crux of our problem in the previous two sections.
Now, let us perform a systematic study of the solutions to the six constraints for each of the
allowed bases for the elliptic fibration.
Enriques base spaces have been shown to be ruled out by effectiveness (see Section 6.1 of
ref.~\cite{Donagi:2004ia}). This leaves us with only three possible choices~:
Hirzebruch surfaces, their blow-ups, and del Pezzo surfaces.
We will address the spectral cover bundles on them case by case.

\subsection{Hirzebruch Surfaces}\label{s:Fr}
We begin with the Hirzebruch surfaces $\mathbb{F}_r$, which are $\mathbb{P}^1$ fibrations over $\mathbb{P}^1$.
We denote the class of the base $\mathbb{P}^1$ by $S$ and that of the fiber by $E$.  These classes have the following intersection numbers
\beq\label{hirz-inter}
E \cdot E=0,\quad S \cdot E=1,\quad S \cdot S=-r\; .
\eeq
The self-intersection number $r$ is an integer between 0 and 12 where the upper bound comes from a theorem in ref.~\cite{Gross:1994}.\footnote{We thank Mark Gross for pointing this out to us.}  Therefore, there are only 13 Hirzebruch surfaces to consider; we note that this has not been thus far stressed in the literature.

The curves in $\mathbb{F}_r$ live in $H_2(B; \IZ)$, which is in fact spanned by $S$ and $E$.
Moreover, every effective curve can be expressed as a linear combination of these generators with non-negative integer coefficients, so we express our effective curve $\eta$ as
\beq
\eta = aS + bE \ , \qquad a,b \in \IZ_{\ge 0} \ .
\eeq
Constraint 1.) requires that the linear system $\vert\eta\vert$ be base-point free in $\mathbb{F}_r$.
This is the case if $\eta\cdot S\ge 0$ (see ref.~\cite{Donagi:2004ia}) or, in  terms of the coefficients,
\begin{enumerate}
\item[1.)] \quad $b \ge r \ a \ .$
\end{enumerate}

To compute the other constraints we need the Chern classes of $\mathbb{F}_r$ which are given by
\bea
c_1(\mathbb{F}_r) &=& 2S + (r+2)E,\\
c_2 (\mathbb{F}_r) &=& 4.
\eea
Combining our above condition for effectiveness, constraints 2.) and 3.) become
\begin{enumerate}
	\item[2.)]\quad $a \ge 2n, \quad b \ge n(r+2)$ \ ,
	\item[3.)]\quad $a \le 24, \quad b \le 12(r+2)$ \ .
\end{enumerate}
Already, from these two conditions we see that the coefficients $a,b$ are bounded and can only have a finite number of solutions for $\eta$.

Next, constraint 4.) becomes
\begin{enumerate}
	\item[4.)]\quad $92 +\frac{n^3-n}{3}-\frac{n}{2}(\lambda^2-\frac{1}{4}) \left(2ab-2na-2nb+nra-ra^2\right) \ge  \left\{
\begin{array}{cl}
	0 & \text{if } r<3, \\
	96+a r-2a-2b & \text{if } r\ge 3,
\end{array} \right. $
\end{enumerate}
The first case, for $r<3$, corresponds to the standard situation, discussed in Section 2.3, where a class $W=W_B+a_f F$ in $X$ is effective iff $W_B$ is effective and $a_f\geq 0$. However, for Hirzebruch surfaces $\mathbb{F}_r$ with $r\geq 3$, the condition $a_f\geq 0$ is replaced by
\beq
  a_f\geq 96+ar-2a-2b,
 \eeq
 and this leads to the more complicated constraint for this case.  In any event, the above condition 4.) becomes a quadratic inequality for $\lambda$ which leads to a finite number solutions.

Finally, constraint 5.) amounts to
\begin{enumerate}
	\item[5.)] \quad $N_{\textrm {gen}}'=\vert \lambda \left(2ab-2na-2nb+nra-ra^2\right) \vert = 3k$,
\end{enumerate}
and from the above expression for the Chern classes and the intersection numbers we have that $c_1(\mathbb{F}_r)^2 = 8$ (interestingly, both this and $c_2(\mathbb{F}_r)$
are independent of $r$) so that 
the Euler character \eqref{eulerB} for $X$ becomes $-480$. Hence constraint 6.) becomes
\begin{enumerate}
	\item[6.)] \quad $480/k \,\in \mathbb{N}$.
\end{enumerate}

In all, the six constraints have therefore become very concrete inequalities in $a,b,\lambda$, and $k$, given $r=0,\ldots,12$.
Indeed, $\lambda$ is integral or half-integral according to $n$, and $a,b$, and $k$ are positive integers. As discussed earlier, constraints 1.), 2.) and 3.) immediately give a finite number of possibilities for $a$ and $b$ which are simply lattice points in a polygon. Furthermore, condition 4.) restricts the possible values of $\lambda$. Hence, we have indeed a finite number of solutions.
On this set, we can then impose the phenomenological conditions 5.) and 6.). This will typically lead to a large reduction of the number of viable models.

To explicitly solve the six equations is straightforward though tedious. A complete lattice point search is
implemented using Mathematica and C++. We present some illustrative examples
of spectral bundles over some of the surfaces in Table \ref{tab:hirzeg}, and
a tally of all the solutions in Table \ref{tab:hirz}.

The bundles in Table \ref{tab:hirzeg} are, according to Section \ref{s:spec}, specified by the integers $n,a,b$, and the (half-)integer $\lambda$. We see that we can produce quite small numbers of net generations. This should be contrasted with the results in refs.~\cite{Anderson:2007nc,Anderson:2008uw}. One observation is that the smaller Hirzebruch surfaces tend to produce models with fewer generations. Indeed, the minimum possible number of generations achievable for each Hirzebruch $\mathbb{F}_r$ decreases
with $r$.
\begin{table}[h!!!]	\centering
$\begin{array}{|c|c|c|c|c|} \hline
\mbox{Base} & n & (a,b) & \lambda & \mbox{\# generations} \\
	\hline \hline
\mathbb{F}_0 &3 & (7,6) & \frac12 & 3 \\ \hline
\mathbb{F}_1 &5 & (13,15)& \frac12& 3\\ \hline
\mathbb{F}_2 &3 & (6,13)& \frac12&3 \\ \hline
\mathbb{F}_3 &3 & (6,18)& \frac12&9 \\ \hline
\mathbb{F}_3 &4 & (8,26)& 1 & 48 \\ \hline
\end{array}$
\mycaption{Some examples of stable vector bundles on elliptically fibered Calabi-Yau threefolds over the first few Hirzebruch surfaces. 
The bundle is specified by integers $n,a,b$, and the integer or half-integer
$\lambda$. We show examples in which the net number of GUT particle generations is equal to $3k$ for some natural number $k$, and such that $k$ divides the Euler number of the Calabi-Yau threefold. We can think of $k$ as the order of a possible discrete group of symmetries.}
\label{tab:hirzeg}
\end{table}

Table \ref{tab:hirz} gives the number of solutions of $SU(3)$, $SU(4)$, and $SU(5)$ bundles on the elliptic threefolds fibered over each of the Hirzebruch surfaces. 
For comparison we have also included two additional sets of results. The first three column represent solutions to the constraints
1.) to 4.) only. Hence, these are stable bundles satisfying anomaly cancellation but their number of generations is not necessarily a multiple of three.\comment{The coefficient of $\lambda^2$ in constraint 4.) is always non-negative, but for $r=0,1,2$ there are a few cases (a dozen) where it vanishes; this implies that the parameter $\lambda$ is arbitrary and there are an infinity of solutions. We simply neglect those cases.} 
Interestingly, $SU(4)$ bundles are the most rare.
The three middle columns count the number of solutions satisfying all six constraints. This leads to an order 10 reduction in the number of bundles. Finally, in the rightmost three columns, we impose the extra condition that $k \le 10$. This is a reasonable constraint because it is in general difficult to find discrete symmetries of very large order. A further reduction is thus seen.
We also find some solutions (shown in parentheses) which gives exactly three generations without the need to quotient by any discrete group, that is $k=1$. These are solutions which correspond to three-generation Grand Unified Theories rather than Standard Model-like theories. These are quite uncommon (only 20 out of the 246), and are concentrated on the first three Hirzebruch surfaces.

\vspace{1cm}

\begin{table}[h]
\centering
\begin{tabular}{|c|c|c|c||c|c|c||c|c|c|} \hline
Constraints & \multicolumn{3}{|c||}{1.) -- 4.)}
& \multicolumn{3}{|c||}{1.) -- 6.)} & \multicolumn{3}{|c|}{1.) -- 6.) and $k\le10$} \\ \hline
Base & $SU(3)$ & $SU(4)$ & $SU(5)$ & $SU(3)$ & $SU(4)$ & $SU(5)$ & $SU(3)$ & $SU(4)$ & $SU(5)$ \\
\hline \hline
$\mathbb{F}_0$ & 756&74&458& 104 &18&34& 48(4)& 6 & 18  \\ \hline
$\mathbb{F}_1$ &878&108 &602& 140&32&58&  56(6)& 20 (4) & 38 (4)   \\ \hline
$\mathbb{F}_2$ &740&40&454&68&10&24& 24 (2) & 4 & 10    \\ \hline
$\mathbb{F}_3$ &666&16&352&66&4&14& 12 & 0 & 2  \\ \hline
$\mathbb{F}_4$ &650 &4&306 &36&2&10& 6 & 0 & 0   \\ \hline
$\mathbb{F}_5$ & 660&0&280 &40&0&2 & 2 & 0 & 0   \\ \hline
$\mathbb{F}_6$ & 682&0&266 &28&0&8 & 0 & 0 &0   \\ \hline
$\mathbb{F}_7$ & 710&0&258 &30&0&8 & 0 & 0 & 0   \\ \hline
$\mathbb{F}_8$ & 740&0&252 &16&0&6 & 0 & 0 & 0   \\ \hline
$\mathbb{F}_9$ & 774&0&250 &24&0&8 & 0 & 0 & 0    \\ \hline
$\mathbb{F}_{10}$ & 810&0& 250&18&0&6 & 0 & 0 & 0    \\ \hline
$\mathbb{F}_{11}$ & 846&0 &250& 22&0&4& 0 & 0 & 0    \\ \hline
$\mathbb{F}_{12}$ &882&0 &250& 18&0&4&  0 & 0 & 0    \\ \hline \hline
\underline{Total} &9794&242 &4228&610 &66&196& 148(12) & 30(4) &68(2)  \\ \hline
\end{tabular}	
\mycaption{The number of stable $SU(n)$ vector bundles from the spectral construction over Calabi-Yau threefolds fibered over the Hirzebruch surfaces and satisfying anomaly cancellation for $n=3,4,5$ (corresponding respectively to gauge groups $E_6$, $SO(10)$, or $SU(5)$) are given in the first three columns. The middle three columns tally those which also give rise to a number of net GUT particle generations divisible by three (that is, they satisfy all our six constraints). The right-most three columns represent the bundles which also require $k$, the order of a possible discrete symmetry, to be less or equal to $10$. The numbers in parentheses indicate models with exactly three net generations where they exist.}\label{tab:hirz}
\end{table}

\begin{figure}[h!!!]
\begin{tabular}{cc}
(a) \includegraphics[width=0.45\textwidth]{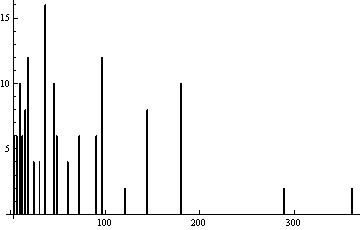} &
(b) \includegraphics[width=0.45\textwidth]{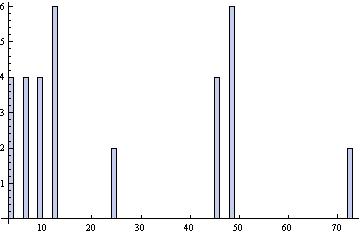} \\
(c) \includegraphics[width=0.45\textwidth]{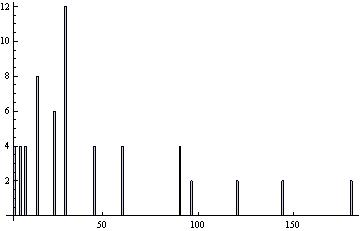} &
(d) \includegraphics[width=0.45\textwidth]{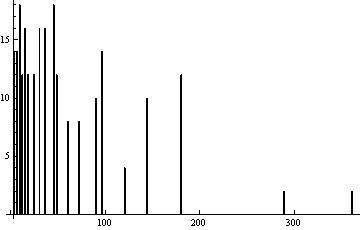}
\end{tabular}
\mycaption{Histograms of the net number of GUT particle generations (such that the number equals $3k$ for some natural number $k$ and such
that $k$ divides the Euler number) for the elliptic Calabi-Yau threefold fibered over
the first Hirzebruch surface $\mathbb{F}_1$ for stable $SU(n)$-bundles at respectively (a) $n=3$, (b) $n=4$, (c) $n=5$, and (d) combined.
The vertical axis is the number of bundles, and the horizontal one the net number of generations.}\label{f:F1}
\end{figure}

\begin{figure}[h!!!]
\begin{tabular}{cc}
(a)\includegraphics[width=0.4\textwidth]{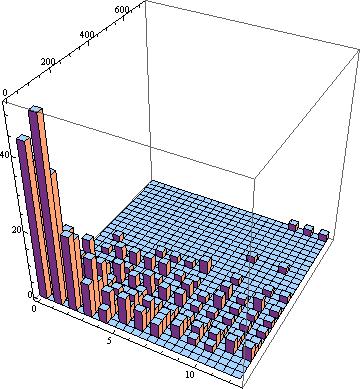} & 
(b)\includegraphics[width=0.4\textwidth]{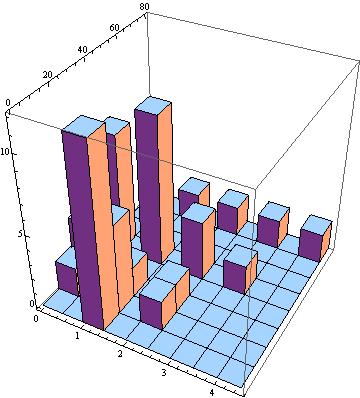} \\
(c)\includegraphics[width=0.4\textwidth]{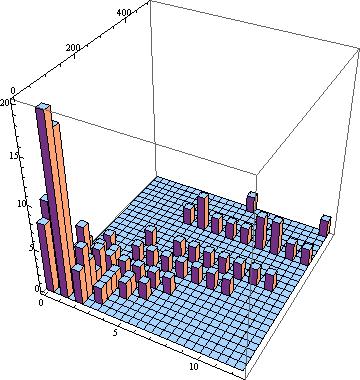} &
(d)\includegraphics[width=0.4\textwidth]{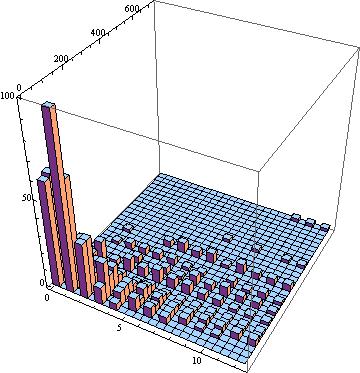} 
\end{tabular}
\mycaption{Histograms of the net number of GUT particle generations (such that the number equals $3k$ for some natural number $k$ and such
that $k$ divides the Euler number) for the elliptic Calabi-Yau threefolds fibered over
all the Hirzebruch surfaces $\mathbb{F}_{0,\ldots,12}$ for stable $SU(n)$-bundles at respectively (a) $n=3$, (b) $n=4$, (c) $n=5$,
and (d) combined. The vertical axis is the number of bundles, and one of the horizontal axes is the net number of generations while the other, from 0 to 12, 
labels the specific Hirzebruch surfaces. Note that, from (b), there are no stable $SU(4)$ bundles with a number of generations divisible by 3 for the fifth and higher Hirzebruch surfaces.
}\label{f:Fr3D}
\end{figure}

\newpage

As a further illustration, let us examine the first Hirzebruch surface, which has a good population of solutions, and is also the only case which admits exactly three generations for all $n=3,4,5$. To illustrate the distribution of the number of net generations, we plot a histogram in Figure \ref{f:F1}. We see that most of the models arise at a small number of generations although models with a large number of generations do exist.

To get an idea of the distribution over the entire family, we plot some three-dimensional histograms in Figure \ref{f:Fr3D}.
As before, on the vertical axis we plot the number of solutions, and  on the horizontal ones we plot the number of generations and the number $r=0,\ldots ,12$ characterizing the Hirzebruch surface.

\newpage

\subsection{Blow-ups of Hirzebruch Surfaces}
Our next family of base surfaces is obtained
by blowing up a point on the curve $E$ of a Hirzebruch surface $\mathbb F_r$ for $r=0,1,2,3$ \cite{Gross:1994}.
Such a blow-up is customarily denoted as $\widehat{\mathbb{F}}_r$.

The second homology is easy to describe. In addition to $E$ and $S$ described in eq.~\eqref{hirz-inter},
there is now a new exceptional class $G$, corresponding to the blow-up. Now, 
if we define $F+G=E$, the intersection numbers are given by \cite{hart}
\bea
& E \cdot E=0, \quad S \cdot S=-r,  \quad S \cdot E=1, &  \\
& F \cdot F = G \cdot G =-1, \quad S\cdot F=1, \quad S\cdot G=0, \quad G\cdot F = 1. &
\eea

Here, an effective curve can be expressed as
\beq
\eta = aS + bF + cG,
\eeq
with $a,b,c \in \IZ_{\ge 0}$ but not all 0, and the Chern classes are given by\cite{Grassi:2000fk}
\bea
c_1(\widehat{\mathbb{F}}_r) &=& 2S + (r+2)F + (r+1)G, \\
c_2 (\widehat{\mathbb{F}}_r) &=& 5.
\eea
\comment{
In addition to constraints 1 and 2, the stability of the associated vector bundle $V$ requires that the curve $\eta$ be irreducible 
This implies the conditions
}
The base-point freeness condition can now be guaranteed by \footnote{We are grateful to Antonella Grassi for pointing this out to us.}
\beq
b \ge a,\quad b \ge c, \quad c \ge b-a.
\eeq
The remaining conditions are straightforward and so
without much ado, we can explicitly summarize the six constraints as
\begin{enumerate}
\item[1.)]\quad $b\ge ra$ \quad
and \quad $b \ge a,\quad b \ge c, \quad c \ge b-a$,
\item[2.)]\quad $a\ge 2n, \quad b\ge (r+2)n, \quad c \ge (r+1)n$,
\item[3.)]\quad $a \le 24, \quad b \le 12(r+2), \quad c \le 12(r+1)$,
\item[4.)]\quad $82 + \frac{7}{24}(n^3-n) - \frac{n}{2}(\lambda^2-\frac 1 4)(-ra^2+2ab-b^2+2bc-c^2+(r-2)na-nb-nc) \ge 0$,
\item[5.)]\quad $\vert\lambda (-ra^2+2ab-b^2+2bc-c^2+(r-2)na-nb-nc)\vert = 3k$,
\item[6.)]\quad $420/k \,\in \mathbb{N}$.
\end{enumerate}

We proceed as before in solving these Diophantine inequalities by lattice-point search and tally
the solutions in Table \ref{tab:blow}. Again, we see that $SU(4)$-bundles are the most rare.
As before we show three data sets, those satisfying only constraints 1.) to 4.)
\comment{As for the Hirzebruch surfaces, we get rid of the few cases where the coefficient of $\lambda^2$ in constraint 4.) vanishes.},
those satisfying all six constraints, and those satisfying all six constraints and in addition having the order of the symmetry group $k \le 10$. A dramatic reduction is seen in the number of solutions with the imposition of these constraints. 

\begin{table}[!h]
   \centering
\begin{tabular}{|c|c|c|c||c|c|c||c|c|c|} \hline
  & \multicolumn{3}{|c||}{1.) -- 4.)}
& \multicolumn{3}{|c||}{1.) -- 6.)} & \multicolumn{3}{|c|}{1.) -- 6.) and $k\le10$} \\ \hline
Base & $SU(3)$ & $SU(4)$ & $SU(5)$ & $SU(3)$ & $SU(4)$ & $SU(5)$ & $SU(3)$ & $SU(4)$ & $SU(5)$ \\
\hline \hline
$\widehat{\mathbb{F}}_0$ &2544 &386&1654 &322 &70& 144 & 162 (14) & 48 (8) & 86 (8)  \\ \hline
$\widehat{\mathbb{F}}_1$ &8872 &922&6344 &882 &174 & 500& 384 (32) & 124 (20) & 266 (20)   \\ \hline
$\widehat{\mathbb{F}}_2$ & 6882&512&5008 &530&108&322& 206 (18)  & 80 (12)  &  152 (12)  \\ \hline
$\widehat{\mathbb{F}}_3$ & 5576&148&3504 &328&16&124&  76 (2) & 2 & 20   \\ \hline
\underline{Total} &23874&1968&16510&1062 & 368 & 1090 & 828 (66) & 254 (40)  & 504 (40) \\ \hline
\end{tabular}
\mycaption{
The number of stable $SU(n)$ vector bundles from the spectral construction over Calabi-Yau threefolds fibered over the blow-ups of Hirzebruch surfaces and satisfying anomaly cancellation for $n=3,4,5$ (corresponding respectively to gauge groups $E_6$, $SO(10)$, or $SU(5)$) are given in the first three columns. The middle three columns tally those which also give rise to a number of net GUT particle generations divisible by three (that is, they satisfy all our six constraints). The right-most three columns represent the bundles which also require $k$, the order of a possible discrete symmetry, to be less or equal to $10$.
}
\label{tab:blow}
\end{table}

\subsection{Del Pezzo Surfaces}
We are finally left with the del Pezzo family of surfaces. It will turn out that, because some higher members of this family have a large number of generators for the Mori cone, these give rise to the most number of bundles. Indeed, elliptic fibrations over specific del Pezzo surfaces have been favorable in constructing realistic models for the past few years \cite{Braun:2005ux,Bouchard:2005ag}.

Let us begin by introducing the geometry.
The del Pezzo surfaces $d\mathbb{P}_r$ is given by $\mathbb{P}^2$, the complex projective plane, blown-up at $r$ generic points. There are only ten del Pezzo surfaces, with $r=0,\ldots,9$, for which elliptic fibration is allowed. The first member,
$d\mathbb{P}_0$, is just $\IP^2$, and the second, $d\mathbb{P}_1$, is isomorphic to the first Hirzebruch surface, $\mathbb{F}_1$.
 
Again, we need the second homology group $H_2(d\mathbb{P}_r,\mathbb{Z})$ to describe the curve classes. The generators are easy to obtain~: they are simply the hyperplane class $l$ in $\mathbb{P}^2$ as well as the exceptional blow-up divisors  $E_i$ with $i=1, \ldots, r$.
They have the following intersection numbers (see for example ref.~\cite{Donagi:2004ia})~:
\beq
l \cdot l=1,\quad l \cdot E_i=0,\quad E_i \cdot E_j=-\delta_{ij}.
\eeq
The Chern classes are given by
\bea
c_1(d\mathbb{P}_r) &=& 3l - \sum_{i=1}^r E_i,\\
c_2 (d\mathbb{P}_r) &=& 3+r.
\eea

The Mori cone for the del Pezzo surfaces is not as simple as the one for the previous cases; its generators are listed in Table \ref{tab:MoridP}. Every effective class can be written as linear combinations of these generators with non-negative integer coefficients.
\begin{table}[t]
	\centering
		\begin{tabular}{|c|c|c|} \hline
		$r$ & Generators \qquad\scriptsize{($i<j<\ldots \le r$)} & Number \\
		\hline \hline
		0 & $l$ & 1 \\ \hline
		1 & $E_1, l-E_1$ & 2 \\ \hline
		2 & $E_i, l-E_i-E_j$ & 3 \\ \hline
		3 & $E_i, l-E_i-E_j$ & 6 \\ \hline
		4 & $E_i, l-E_i-E_j$ & 10 \\ \hline
		5 & $E_i, l-E_i-E_j, 2l-E_i-E_j-E_k-E_l-E_m$ & 16 \\ \hline
		6 & $E_i, l-E_i-E_j, 2l-E_i-E_j-E_k-E_l-E_m$ & 27 \\ \hline
		7 & $E_i, l-E_i-E_j, 2l-E_i-E_j-E_k-E_l-E_m$,& \\
		  & $3l-2E_i-E_j-E_k-E_l-E_m-E_n-E_o$ & 56 \\ \hline
		8 & $E_i, l-E_i-E_j, 2l-E_i-E_j-E_k-E_l-E_m$, &\\
		  & $3l-2E_i-E_j-E_k-E_l-E_m-E_n-E_o, \ldots$ & 240 \\ \hline
		9 & \ldots & $\infty$ \\ \hline
		\end{tabular}
	\mycaption{Generators of the Mori cone (of effective curves) for del Pezzo surfaces $d\mathbb{P}_r$.}
	\label{tab:MoridP}
\end{table}
It is a concern that $d\mathbb{P}_9$ has an infinite dimensional Mori cone. This may contradict our finiteness result. Luckily, as discussed in Section 6.2 of ref.~\cite{Donagi:2004ia}, the {\em generic} $d\mathbb{P}_9$ surface is ruled out by the requirement of a non-zero number of generations, eq.~\eqref{cond4.1}. The basic reason is that this surface is itself an elliptic fibration over $\IP^1$ with fiber class $f$. From effectiveness, $\eta$ must be proportional to $f$ and as a result the number of generations is zero. Therefore we need not consider this surface. 

However, we must point out that {\em special} $d\mathbb{P}_9$ surfaces, where additional isometries are found in special points of moduli space, are allowed.  Indeed, all the successful models in the literature based on this surface are special $d\mathbb{P}_9$ \cite{Braun:2005ux,Bouchard:2005ag}. We will not consider these special cases here.

Finally, we need conditions for the linear system $|\eta |$ to be base-point free. On $d\mathbb{P}_r$ for $2\le r\le 7$, this is the case if the divisor $\eta$ is such that $\eta\cdot E \ge 0$ for every   
curve $E$ which satisfies the two properties $E\cdot E = -1$ and $E\cdot c_1(d\mathbb{P}_r)=1$. Therefore, this condition is again  translated into constraints on intersection numbers.

Now we are ready to write our six constraints in terms of coefficients of $\eta$ expanded into the Mori cone, as well as $\lambda$ and $k$.
Indeed, from Table \ref{tab:MoridP} we see that there is a very rapidly increasing number of generators as $r$ increases. Thus we have increasing numbers of coefficients to deal with and this complicates our algorithmic computations. We treat each case of $r$ separately.

\begin{table}[t]
{\hspace{-2cm}
\begin{tabular}{|c|c|c|c||c|c|c||c|c|c|} \hline
& \multicolumn{3}{|c||}{1.) -- 4.)}
& \multicolumn{3}{|c||}{1.) -- 6.)} & \multicolumn{3}{|c|}{1.) -- 6.) and $k\le10$} \\ \hline
Base & $SU(3)$ & $SU(4)$ & $SU(5)$ & $SU(3)$ & $SU(4)$ & $SU(5)$ & $SU(3)$ & $SU(4)$ & $SU(5)$ \\
\hline \hline
$d\mathbb{P}_0$ & 62 & 10 & 44 &12&2&6 & 4 & 0 & 2 \\ \hline
$d\mathbb{P}_1$ &878&108 &602& 140&32&58&  56(6)& 20 (4) & 38 (4)   \\ \hline
$d\mathbb{P}_2$ & 8,872 & 922 & 6,344 & 882  & 174  & 500 & 384(32) &124(20)&266 (20) \\ \hline
$d\mathbb{P}_3$ & 4,564,124 & 399,446 & 6,080,464 & 291,282 & 82,622 & 529,102  & $\begin{array}{l} 101,204 \\(5,306)\end{array}$
& $\begin{array}{l} 62,098  \\ (5,198)\end{array}$
& $\begin{array}{l} 247,724  \\ (8,738) \end{array}$
\\ \hline
$d\mathbb{P}_4$ &\ldots & \ldots & \ldots &
       \ldots & \ldots & \ldots & \ldots   & \ldots & \ldots \\ \hline
\end{tabular}
}
\mycaption{
The number of stable $SU(n)$ vector bundles from the spectral construction over Calabi-Yau threefolds fibered over the first four del Pezzo surfaces and satisfying anomaly cancellation for $n=3,4,5$ (corresponding respectively to gauge groups $E_6$, $SO(10)$, or $SU(5)$) are given in the first three columns. The middle three columns tally those which also give rise to a number of net GUT particle generations divisible by three (that is, they satisfy all our six constraints). The right-most three columns represent the bundles which also require $k$, the order of a possible discrete symmetry, to be less or equal to $10$. The numbers in parentheses indicate those, where possible, with exactly three net generations.}
\label{tab:dP}
\end{table}

As an example, let us discuss $d\mathbb{P}_2$ explicitly. There are three generators of the Mori cone and
we can write
\beq
\eta = a E_1 + b(l-E_1-E_2) + c E_2 \ .
\eeq
Subsequently, the six constraints become~:
\begin{enumerate}
\item[1.)]\quad $b \ge a, \quad a+c \ge b, \quad b \ge c$,
\item[2.)]\quad $a \ge 2n, \quad b \ge 3n, \quad c \ge 2n$,
\item[3.)]\quad $a \le 24, \quad b \le 36, \quad c \le 24$,
\item[4.)]\quad $82 +\frac{7}{24}(n^3-n)-\frac{n}{2}(\lambda^2-\frac{1}{4})(-a^2+2ab-b^2+2bc-c^2-na-nb-nc) \ge 0$,
\item[5.)]\quad $N_{gen}'=|\lambda (-a^2+2ab-b^2+2bc-c^2-na-nb-nc)|= 3k$,
\item[6.)]\quad $ 420/k \in \mathbb{N}$.
\end{enumerate}
Again, we can find all solutions via an exhaustive lattice-point search.

\begin{figure}[t]
\begin{tabular}{cc}
(a) \includegraphics[width=0.45\textwidth]{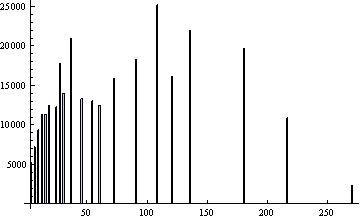} &
(b) \includegraphics[width=0.45\textwidth]{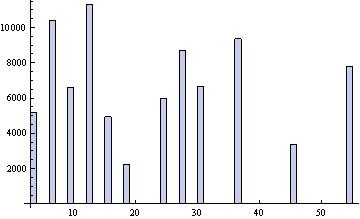} \\
(c) \includegraphics[width=0.45\textwidth]{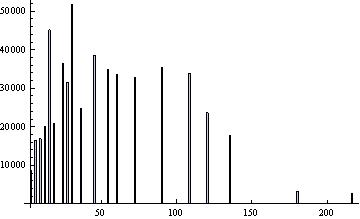} &
(d) \includegraphics[width=0.45\textwidth]{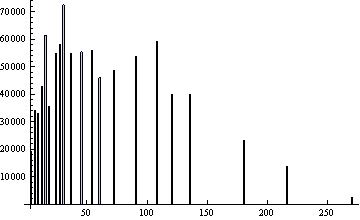}
\end{tabular}
\mycaption{Histograms of the net number of grand unified particle generations (such that the number equals $3k$ for some natural number $k$ and such
that $k$ divides the Euler number) for the elliptic Calabi-Yau threefold fibered over
the third del Pezzo surface $d\IP_3$ for stable $SU(n)$-bundles at respectively (a) $n=3$, (b) $n=4$, (c) $n=5$, and (d) combined.
The vertical axis is the number of bundles, and the horizontal one the net number of generations.}\label{f:dP3}
\end{figure}

In Table \ref{tab:dP} we record the tally of solutions for the first four del Pezzo surfaces.
We see that the number of solutions grow exponentially, and $d\mathbb{P}_r$ for $4\leq r\leq 8$ exceeds
what we can perform with present computer power.

We see that there is an enormous number of stable bundles which satisfy our physical constraints. As an illustration, let us plot the results for $d\mathbb{P}_3$, the richest so far, in Figure \ref{f:dP3}.

\section{Conclusion and Prospects}

In this paper, inspired by recent advances in applying computer algebra and
computational algebraic geometry to string phenomenology \cite{Anderson:2007nc,fluxcomp}, we
initiated the construction and statistical analysis of {\em the largest set of explicit stable bundles to date.}
We carried out a classification of spectral cover vector bundles, compatible with heterotic model-building constraints, over elliptically fibered Calabi-Yau manifolds. For both Hirzebruch and blown-up Hirzebruch base spaces we obtained a complete classification of anomaly-free bundles with about $30,000$ $SU(3)$ cases, $20,000$ $SU(5)$ cases, and only about $2,000$ $SU(4)$ cases. It is interesting to note the difference in numbers between $SU(4)$ and the other two structure groups; this is related to a case distinction for the parameter $\lambda$ in the spectral cover construction. 

We then imposed two physical constraints on these bundles, namely the three-generation constraint and the  requirement that the order of a possible discrete symmetry group is at most $10$. This led to a dramatic reduction in the number of viable models to about $1,700$, most of them concentrated on blown-up Hirzebruch base spaces.

For del Pezzo base spaces the situation was somewhat different due to the large number of Mori cone generators for the higher del Pezzo surfaces. In practice, we could only perform a complete classification up to $d\mathbb{P}_3$, where we found {\em over $11$ million anomaly-free stable bundles.} 
As before, the number of $SU(4)$ bundles is relatively small with about $400,000$ cases. 
We should stress that among all these bundles the number of those that have different second and third Chern classes is smaller by a factor 100 approximately.
They could still however be genuinely different bundles, given that they have different coefficients for the expansion of $\eta$ in terms of Mori cone generators or different values of $\lambda$.

Imposing the physical constraints led to a reduction of the number of models by a factor of more than $10$ but we are still left with about $400,000$ viable models at this stage. 
We did not explicitly classify bundles on del Pezzo surfaces $d\mathbb{P}_r$ with $r>3$, as this task exceeds current computer power. 

For Hirzebruch base spaces our approach led to a relatively small number of about $1,700$ viable models which can now be studied in detail. For del Pezzo base spaces, on the other hand, we clearly need more physical constraints which can be systematically imposed in order to filter out the presumably small number of physically promising examples. A systematic search for discrete symmetries of elliptically fibered Calabi-Yau manifolds and spectral cover bundles over them would likely lead to very tight constraints, but performing such a search is a very challenging tasks indeed. 

In addition, by methods similar to the one employed in this paper, one could classify spectral cover bundles with $U(n)$ structure groups over elliptically fibered Calabi-Yau manifolds. Discrete symmetries and Wilson lines are not required for such models and systematically imposing detailed physical constraints might be a more straightforward task. These issues are currently under investigation.

\section*{Acknowledgments}
The authors would like to express their sincere gratitude to Antonella Grassi, Mark Gross and Tony Pantev for many helpful discussions. M.~G.~thanks the Berrow Foundation for supporting his work. 
Y.-H. H is indebted to the UK STFC for
an Advanced Fellowship as well as the FitzJames Fellowship of Merton College, Oxford. A. L. is
supported by the EC 6th Framework Programme MRTN-CT-2004-503369.


\end{document}